\documentclass[aps,pra,twocolumn,nofootinbib,10pt]{revtex4-1}

\usepackage{natbib}
\usepackage{amsmath,amssymb,bm}
\usepackage{dsfont}
\usepackage{graphics}

\newcommand{\tr}{\text{tr}}
\newcommand{\eqtri}{\triangleq}

\newcommand{\ket}[1]{\left|{#1}\right\rangle}
\newcommand{\bra}[1]{\left\langle{#1}\right|}

\newcommand{\Dcirc}{{\cal D}^{\circ}}

\begin{document}

\title{Non-Markovian evolution of multiphoton states in turbulence}
\author{Filippus S. \surname{Roux}}
\email{stefroux@gmail.com}
\affiliation{University of Kwazulu-Natal, Private Bag X54001, Durban 4000, South Africa}

\begin{abstract}
An evolution equation for multiphoton states propagating through turbulence is derived without making a Markovian approximation. The state is represented as a Wigner functional to incorporate all spatiotemporal degrees of freedom. The resulting non-Markovian evolution equation is used to argue that initial Gaussian states do not remain Gaussian during propagation. Possible solutions of this evolution equation are discussed.
\end{abstract}

\maketitle

\section{\label{intro}Introduction}

Multiphoton quantum states provide benefits in a variety of applications, such as quantum information processing and quantum metrology. An understanding of the propagation of multiphoton quantum states through turbulence is necessary for the implementation of quantum cryptography and continuous variable teleportation in free-space quantum communication systems \cite{cvteleport,wig1,semenov,fstelez,heim,cvcomrev,advtelport,cvqkd,derkach,cvtelml}.

The effect of a turbulent medium on single- or bi-photon states has been studied extensively \cite{paterson,sr,qturb4,qturb3,ipe,turbsim,leonhard,notrunc} and quantum communication systems based on such states have been physically demonstrated \cite{malik,vallone,krenn2,qkdturb}. On the other hand, the effect of a turbulent medium on multiphoton states has received less attention, being a significantly more complex problem. One way in which the effect of turbulence on multiphoton states has been considered, is to model it with a loss mechanism \cite{semenov}, effecting only the photon-number degrees of freedom of the state, while ignoring its effect on the other degrees of freedom. Using a Wigner functional approach, an evolution equation for a multiphoton state propagating through turbulence under the Markovian assumption has been developed previously by the current author \cite{ipfe}.

Here, we consider the non-Markovian case and provide a simpler derivation for the evolution equation, compared to the derivation in \cite{ipfe}. The resulting equation has the same form obtained in \cite{ipfe}, but with more complicated expression for the vertex kernels.

Based on the nature of the dissipative terms in the evolution equation, we argue that an initial Gaussian state loses its Gaussian nature during propagation through turbulence. In other words, the full non-Markovian evolution equation does not have Gaussian solutions. Without the dissipative terms, a transformation of the argment of the Wigner functional of the state would suffice as a solution of the evolution equation. Thermal states can be approximated as Gaussian state solutions, by considering their kernels as the second moments of the Wigner functionals. We also consider the state based on the loss model \cite{semenov} as a possible solution, but find that it does not solve the evolution equation.

\section{Derivation}

\subsection{Classical equation of motion}

To derive the evolution equation, we start with the equation of motion for paraxial propagation of classical light through turbulence. It is given by
\begin{equation}
\nabla_T^2 g(\mathbf{X},z) - i 2k\partial_z g(\mathbf{X},z) + 2k^2 \tilde{n}(\mathbf{X},z) g(\mathbf{X},z) = 0 ,
\end{equation}
where $\nabla_T^2=\partial_x^2+\partial_y^2$ is the transverse Laplacian, $g(\mathbf{X},z)$ is the slow varying part of a scalar electromagnetic phasor field, $\mathbf{X}$ is the two-dimensional transverse coordinate vector, $z$ is the propagation distance, $k=2\pi/\lambda$ is the wavenumber (which implies a monochromatic approximation), and $\tilde{n}(\mathbf{X},z)$ is the fluctuation in the refractive index of the atmosphere, so that the refractive index is represented as $n=1+\tilde{n}$. The scalar electromagnetic field is represented by an inverse Fourier transform with a $z$-dependent angular spectrum
\begin{equation}
g(\mathbf{X},z) = \int G(\mathbf{K},z) \exp\left(-i \mathbf{K}\cdot\mathbf{X}\right)\ \frac{d^2k}{(2\pi)^2} ,
\end{equation}
where $\mathbf{K}$ is the two-dimensional transverse wave vector. In a similar way, the refractive index fluctuation is represented in terms of a $z$-dependent spectrum
\begin{equation}
\tilde{n}(\mathbf{X},z) = \int N(\mathbf{K},z) \exp\left(-i \mathbf{K}\cdot\mathbf{X}\right)\ \frac{d^2k}{(2\pi)^2} .
\end{equation}
Since the refractive index fluctuation is a real-valued function, $N^*(\mathbf{K},z)=N(-\mathbf{K},z)$.

Substituting these inverse Fourier transforms into the classical equation of motion and performing a Fourier transform on the result, we obtain
\begin{align}
\partial_z G(\mathbf{K},z) = & \frac{i |\mathbf{K}|^2}{2k} G(\mathbf{K},z) \nonumber \\
& -i k \int N(\mathbf{K}-\mathbf{K}',z) G(\mathbf{K}',z)\ \frac{d^2k'}{(2\pi)^2} .
\end{align}
The first term on the right-hand side is a free-space propagation terms. It can be removed when we define
\begin{equation}
G(\mathbf{K},z) = \exp\left(\frac{i z|\mathbf{K}|^2}{2k}\right) G_c(\mathbf{K},z) .
\label{paropl}
\end{equation}
The equation of motion for $G_c(\mathbf{K},z)$ is
\begin{align}
\partial_z G_c(\mathbf{K},z) = & -i k \int \exp\left[\frac{i z}{2k}\left(|\mathbf{K}'|^2-|\mathbf{K}|^2\right)\right] \nonumber \\
& \times N(\mathbf{K}-\mathbf{K}',z) G_c(\mathbf{K}',z)\ \frac{d^2k'}{(2\pi)^2} .
\end{align}

\subsection{Quantum equation of motion}

The classical equation of motion is quantized by replacing the angular spectrum by a Fourier domain field operator. The latter is obtained from the transverse Fourier transform of the quantized electric field under the paraxial approximation. The scalar electric field operator $\hat{E} = \vec{\eta}_0\cdot\hat{\mathbf{E}}$ is extracted with a suitable state of polarization. The result of the Fourier transform, applied to the annihilating part only, is given by
\begin{align}
\hat{G}(\mathbf{K},z) = & \int \hat{E}^{+}(\mathbf{x}) \exp\left(i \mathbf{K}\cdot\mathbf{X} \right)\ d^2x \nonumber \\
= & -i \sqrt{\frac{\hbar k}{2\epsilon_0}} \hat{a}(\mathbf{K}) \exp\left(i \frac{z}{2 k}|\mathbf{K}|^2\right) ,
\label{gopdef}
\end{align}

In a dielectric medium, the permittivity would be $\epsilon$. However, since the fluctuations are dealt with in terms of the dynamics under investigation and the average refractive index of air is approximately 1, we use $\epsilon_0$ here.

In general, the Fourier domain field operators depend on $\mathbf{K}$ and the angular frequency $\omega$ (the {\em optical beam variables}). Since a turbulent medium is a linear system, we can focus on the monochromatic case and ignore the $\omega$-dependence. The monochromatic assumption is a prerequisite for the paraxial approximation.

The quadratic phase factor in Eq.~(\ref{gopdef}) is a result of the paraxial approximation and represents a free-space propagation phase factor. We can absorb it into the definitions of the field operators, in analogy to the partial solution of the classical scalar field in Eq.~(\ref{paropl}). The quantized equation of motion is then given by
\begin{align}
\partial_z \hat{G}_c(\mathbf{K},z) = & -i k \int \exp\left[\frac{i z}{2k}\left(|\mathbf{K}'|^2-|\mathbf{K}|^2\right)\right] \nonumber \\
& \times N(\mathbf{K}-\mathbf{K}',z) \hat{G}_c(\mathbf{K}',z)\ \frac{d^2k'}{(2\pi)^2} ,
\label{eomq}
\end{align}
where we defined the {\em co-propagating} field operator as
\begin{align}
\hat{G}_c(\mathbf{K},z) = & \exp\left(\frac{-i z|\mathbf{K}|^2}{2k}\right) \hat{G}(\mathbf{K},z) \nonumber \\
= & -i \sqrt{\frac{\hbar k}{2\epsilon_0}} \hat{a}(\mathbf{K}) .
\end{align}
The commutation relation for the scalar field operators
\begin{equation}
[\hat{G}_c(\mathbf{K},z),\hat{G}_c^{\dag}(\mathbf{K}',z)] = (2\pi)^2 \frac{\hbar k}{2\epsilon_0} \delta(\mathbf{K}-\mathbf{K}') ,
\end{equation}
follow from those for the ladder operators
\begin{equation}
\left[ \hat{a}(\mathbf{K}_1), \hat{a}^{\dag}(\mathbf{K}_2) \right] = (2\pi)^2 \delta(\mathbf{K}_1-\mathbf{K}_2) .
\label{kommparax}
\end{equation}

\subsection{Propagation operator}

The evolution equation of the quantum field operator (in the Heisenberg picture) is of the form
\begin{equation}
i \hbar\frac{d}{dz} \hat{G}_c(z) = [\hat{G}_c(z),\hat{P}] ,
\label{hipe}
\end{equation}
where $\hat{P}$ represents the propagation operator in the presence of scintillation. The ansatz for this propagation operator in the co-propagating frame has the form
\begin{align}
\hat{P} = & \int \hat{G}_c^{\dag}(\mathbf{K},z) M(\mathbf{K},\mathbf{K}',z)
\hat{G}_c(\mathbf{K}',z)\ \frac{d^2 k}{(2\pi)^2}\ \frac{d^2 k'}{(2\pi)^2} \nonumber \\
\eqtri &\ \hat{G}_c^{\dag}\diamond M\diamond\hat{G}_c ,
\end{align}
where $M(\mathbf{K},\mathbf{K}',z)$ is a Hermitian kernel, which means that $M(\mathbf{K}_1,\mathbf{K}_2,z) = M^*(\mathbf{K}_2,\mathbf{K}_1,z)$. We also define a $\diamond$-contraction notation to simplify the expression. After evaluating the commutation in Eq.~(\ref{hipe}), we get
\begin{equation}
i \hbar\frac{d}{dz} \hat{G}_c
= \frac{\hbar k}{2\epsilon_0} \int M(\mathbf{K},\mathbf{K}',z)\hat{G}_c(\mathbf{K}'z)\ \frac{d^2 k'}{(2\pi)^2} .
\end{equation}
Comparing it to the expression of the quantized equation of motion in Eq.~(\ref{eomq}) multiplied by $i \hbar$, we obtain an expression for the kernel in the ansatz, given by
\begin{align}
M(\mathbf{K},\mathbf{K}',z) = & 2\epsilon_0 N(\mathbf{K}-\mathbf{K}',z) \nonumber \\
& \times \exp\left[\frac{i z}{2k}\left(|\mathbf{K}'|^2-|\mathbf{K}|^2\right)\right] .
\label{defm}
\end{align}

\subsection{Wigner functional}

The derivation of the evolution equation in terms of Wigner functionals follows an approach that has been used previous for the evolution of a state in nonlinear media \cite{nosemi}. We use a coherent-state-assisted approach to compute the Wigner functional for the propagation operator. For this purpose, the operator is overlapped by two different coherent states. When the field operator is applied to a coherent state, it produces
\begin{align}
\hat{G}_c(\mathbf{K}) \ket{\alpha}
= & -i \sqrt{\frac{\hbar k}{2\epsilon_0}} \hat{a}(\mathbf{K}) \ket{\alpha} \nonumber \\
= & -i \ket{\alpha} \sqrt{\frac{\hbar k}{2\epsilon_0}} \alpha(\mathbf{K}) ,
\label{kwantg}
\end{align}
where, $\alpha(\mathbf{K})$ is the monochromatic angular spectrum of the coherent state's parameter function. Since we are moving to the Schr{\"o}dinger picture, the operators and the spectral function $\alpha$ lose their $z$-dependences. The kernel $M(\mathbf{K},\mathbf{K}',z)$ retains its $z$-dependences because it represents the inhomogeneous medium.

The overlap of the propagation operator by coherent states on both sides then leads to
\begin{align}
\bra{\alpha_1}\hat{P}\ket{\alpha_2}
= & \hbar \exp\left(-\tfrac{1}{2}\|\alpha_1\|^2-\tfrac{1}{2}\|\alpha_2\|^2+\alpha_1^*\diamond\alpha_2\right) \nonumber \\
& \times \alpha_1^*\diamond M_0 \diamond\alpha_2 ,
\label{apan0}
\end{align}
where
\begin{align}
M_0(\mathbf{K},\mathbf{K}',z) = & k N(\mathbf{K}-\mathbf{K}',z) \nonumber \\
& \times \exp\left[\frac{i z}{2k}\left(|\mathbf{K}'|^2-|\mathbf{K}|^2\right)\right] .
\label{defm0}
\end{align}

Now, we represent the overlapped propagation operator in terms of a generating functional and a construction process. The generating functional is
\begin{align}
\mathcal{G} = & \exp\left(-\tfrac{1}{2}\|\alpha_1\|^2-\tfrac{1}{2}\|\alpha_2\|^2 +\alpha_1^*\diamond\alpha_2\right) \nonumber \\
& \times \exp\left(\alpha_1^*\diamond\mu+\nu^*\diamond\alpha_2 \right) ,
\end{align}
where $\mu$ and $\nu^*$ are auxiliary field variables. The construction operation consists of functional derivatives
\begin{equation}
\mathcal{C} = \hbar\delta_{\mu}\diamond M_0\diamond\ \delta_{\nu}^* ,
\label{konstrukn}
\end{equation}
where
\begin{equation}
\delta_{\mu}(\mathbf{K}) = \frac{\delta}{\delta\mu(\mathbf{K})} ~~~ \text{and} ~~~
\delta_{\nu}^*(\mathbf{K}) = \frac{\delta}{\delta\nu^*(\mathbf{K})} .
\end{equation}
After, applying the functional derivatives of the construction operation on the generating functional, we set the auxiliary field variables to zero.

The generating functional is substituted into the functional integral for the coherent-state-assisted approach to obtain a generating functional for the Wigner functional of the propagation operator. It reads
\begin{align}
\mathcal{W}_{\mathcal{G}}
= & \mathcal{N}_0 \int \exp\left(-2\|\alpha\|^2+2\alpha^*\diamond\alpha_1+2\alpha_2^*\diamond\alpha \right. \nonumber \\
& -\alpha_2^*\diamond\alpha_1-\|\alpha_1\|^2-\|\alpha_2\|^2+\alpha_1^*\diamond\alpha_2 \nonumber \\
& \left. +\alpha_1^*\diamond\mu+\nu^*\diamond\alpha_2\right)\ \Dcirc[\alpha_1,\alpha_2] \nonumber \\
= & \exp\left(\nu^*\diamond\alpha+\alpha^*\diamond\mu-\tfrac{1}{2}\nu^*\diamond\mu \right) ,
\label{wigpropgen}
\end{align}
where $\alpha$ now serves as an {\em integration field variable}, and the functional integration measure is $\Dcirc[\alpha]=\mathcal{D}[\alpha/2\pi]$.

The Wigner functional for the propagation operator can now be obtained by computing
\begin{equation}
W_{\hat{P}}[\alpha] = \left. \mathcal{C} \{ \mathcal{W}_{\mathcal{G}} \} \right|_{\mu=\nu^*=0} .
\end{equation}
However, it is more convenient to postpone the application of the construction operation till after substituting the generating functional into the Wigner functional version of the evolution equation.

\subsection{\label{wigipe}Unitary evolution equation}

Instead of using the Wigner functional for the infinitesimal propagation operator directly in the expression of the evolution equation, we use its representation in terms of the construction operation in Eq.~(\ref{konstrukn}) and the generating functional in Eq.~(\ref{wigpropgen}). The evolution equation in terms of Wigner functionals then reads
\begin{equation}
i \hbar \frac{d}{dz} W_{\hat{\rho}}
= \left. \mathcal{C}\left\{\mathcal{W}_{\mathcal{G}}\star W_{\hat{\rho}}
-W_{\hat{\rho}}\star\mathcal{W}_{\mathcal{G}}\right\} \right|_{\mu=\nu^*=0} ,
\label{ipewig0}
\end{equation}
where $\star$ is the Moyal star product. The calculation of the star products produce
\begin{align}
\begin{split}
\mathcal{W}_{\mathcal{G}}\star W_{\hat{\rho}}
= & \exp\left(\nu^*\diamond\alpha+\alpha^*\diamond\mu-\tfrac{1}{2}\nu^*\diamond\mu \right) \\
& \times W_{\hat{\rho}}\left[\alpha^*+\tfrac{1}{2}\nu^*,\alpha-\tfrac{1}{2}\mu\right] , \\
W_{\hat{\rho}}\star\mathcal{W}_{\mathcal{G}}
= & \exp\left(\nu^*\diamond\alpha+\alpha^*\diamond\mu-\tfrac{1}{2}\nu^*\diamond\mu \right) \\
& \times W_{\hat{\rho}}\left[\alpha^*-\tfrac{1}{2}\nu^*,\alpha+\tfrac{1}{2}\mu\right] .
\end{split}
\label{sterprods}
\end{align}

When we apply the construction operation to the two star products, it leads to the evolution equation
\begin{equation}
i \partial_z W_{\hat{\rho}}
= \alpha^*\diamond M_0\diamond\left(\delta_{\alpha}^* W_{\hat{\rho}}\right)
- \left(\delta_{\alpha} W_{\hat{\rho}}\right)\diamond M_0\diamond\alpha ,
\label{ipewig1}
\end{equation}
where we cancel $\hbar$ on both sides, and defined
\begin{equation}
\delta_{\alpha}(\mathbf{K}) = \frac{\delta}{\delta\alpha(\mathbf{K})} ~~~ \text{and} ~~~
\delta_{\alpha}^*(\mathbf{K}) = \frac{\delta}{\delta\alpha^*(\mathbf{K})} .
\end{equation}
The kernel $M_0$ is defined in Eq.~(\ref{defm0}). We also note that the total derivative becomes a partial derivative because the field variables are independent of $z$.

\subsection{Second order}

The equation in Eq.~(\ref{ipewig1}) represents the first order unitary evolution of the state. This evolution equation is not useful because we don't know the exact bilinear kernel. We only know (some of) its statistical properties. Therefore we can only make predictions about the evolution of the statistical ensemble average of the state. The equation in Eq.~(\ref{ipewig1}) cannot provide the required dynamics when we apply the ensemble averaging, because it is assumed that the refractive index fluctuation has a zero mean, which implies $\langle M_0 \rangle=0$. The result is that, after an ensemble average, we obtain $\partial_z W_{\hat{\rho}}(z)=0$, which corresponds to free-space propagation without turbulence.

To see the effect of the turbulence after an ensemble averaging, we consider the second order. For this purpose, Eq.~(\ref{ipewig1}) is integrated over $z$ so that
\begin{align}
W_{\hat{\rho}}(z)
= & W_{\hat{\rho}}(z_0) - i \int_{z_0}^z \delta_{\alpha}^* W_{\hat{\rho}}(z_1)\diamond M_0^T(z_1)\diamond\alpha^* \nonumber \\
& - \delta_{\alpha} W_{\hat{\rho}}(z_1)\diamond M_0(z_1)\diamond\alpha\ dz_1 .
\label{ipewig1b}
\end{align}
Then we substitute Eq.~(\ref{ipewig1b}) repeatedly back into the first-order equation in Eq.~(\ref{ipewig1}). We perform the back-substitution twice, so that, after the ensemble averages have removed all first order and third order terms, we end up with second order terms only where the $z$-dependences of the Wigner functionals turned into $z_0$-dependences. The functional derivatives are evaluated where possible. The resulting integro-differential equation reads
\begin{widetext}
\begin{align}
\partial_z W_{\hat{\rho}}(z) = & - \int_{z_0}^z
\delta_{\alpha} W_{\hat{\rho}}(z_0)\diamond M_0(z_1)\diamond M_0(z)\diamond\alpha
+\alpha^*\diamond M_0(z)\diamond M_0(z_1)\diamond\delta_{\alpha}^* W_{\hat{\rho}}(z_0) \nonumber \\
& +\alpha^*\diamond M_0(z)\diamond\delta_{\alpha}^* \delta_{\alpha}^* W_{\hat{\rho}}(z_0)\diamond M_0^T(z_1)\diamond\alpha^*
+ \alpha\diamond M_0^T(z)\diamond\delta_{\alpha} \delta_{\alpha} W_{\hat{\rho}}(z_0)\diamond M_0(z_1)\diamond\alpha \nonumber \\
& - \alpha^*\diamond M_0(z)\diamond\delta_{\alpha}^* \delta_{\alpha} W_{\hat{\rho}}(z_0)\diamond M_0(z_1)\diamond\alpha
- \alpha^*\diamond M_0(z_1)\diamond\delta_{\alpha}^* \delta_{\alpha} W_{\hat{\rho}}(z_0)\diamond M_0(z)\diamond\alpha\ dz_1 ,
\label{ipewig2}
\end{align}
\end{widetext}
where we maintain the $\diamond$-contraction notation with the understanding that, for two functional derivatives, the first (second) one is contracted to the left (right).

All terms in Eq.~(\ref{ipewig2}) contain two $M_0$'s. Only one of the $M_0$'s is integrated over $z$. The ensemble average combines the two $M_0$'s in each term into one four-point kernel, as shown below. In the first two terms, two of the legs of the four-point kernel are contracted on each other, turning it into a bilinear kernel. These two terms represent the drift process in the Fokker-Planck equation. By themselves, they allow a solution represented by a transformation of the field variables only and do not represent a change the shape of the initial Wigner functional, apart from a scaling. The remaining four terms represent the diffusion or dissipative process. They can change the shape of the Wigner functional. The four-point kernel resulting from the ensemble average, convert the bilinear terms into terms consisting of four field variables. As a result, these terms tend to destroy the Gaussian nature of the Wigner functional of any initial Gaussian state.

The evolution equation in Eq.~(\ref{ipewig2}) can be considered or interpreted in two different ways. If we consider $z$ as an infinitesimally increased distance beyond $z_0$, then we can convert the equation effectively to a second order differential equation. In that case, $z_0$ is any intermediate position and not the initial position; the equation represents a truly infinitesimal evolution. However, in this case it loses any non-Markovian effects.

To see non-Markovian effects, we treat $z_0$ as the initial position and allow $z$ to take on any value beyond $z_0$, much larger than an infinitesimal propagation. The integral thus remains and we do not convert it to a second order differential equation.

\subsection{\label{nonmark}Ensemble average}

The ensemble averaging process removes all the uneven order terms. The second-order terms contain the ensemble average of the product of two $M_0$'s. Without the $z$-integrations, they are given by
\begin{align}
& \langle M_0(\mathbf{K}_1,\mathbf{K}_2,z_1) M_0(\mathbf{K}_3,\mathbf{K}_4,z_2)\rangle \nonumber \\
= & k^2 \exp\left[\frac{i z_1}{2k}\left(|\mathbf{K}_2|^2-|\mathbf{K}_1|^2\right)
+\frac{i z_2}{2k}\left(|\mathbf{K}_4|^2-|\mathbf{K}_3|^2\right)\right] \nonumber \\
& \times \langle N(\mathbf{K}_1-\mathbf{K}_2,z_1)N(\mathbf{K}_3-\mathbf{K}_4,z_2)\rangle ,
\end{align}
where we substituted in Eq.~(\ref{defm0}). We need to evaluate the ensemble average of the two $N$'s. For this purpose, we use the Fourier transform
\begin{equation}
N(\mathbf{K},z) = \int \tilde{n}(\mathbf{X},z) \exp\left(i \mathbf{K}\cdot\mathbf{X}\right)\ d^2x ,
\end{equation}
and model the refractive index fluctuations as
\begin{equation}
\tilde{n}(\mathbf{x}) = \int \exp\left(-i \mathbf{k}\cdot\mathbf{x}\right) \chi(\mathbf{k})
\left[ \frac{\Phi_n(\mathbf{k})}{\Delta^3} \right]^{1/2} \frac{d^3k}{(2\pi)^3} ,
\label{defn3d}
\end{equation}
where $\Delta$ is a dimension parameter on the frequency domain, $\Phi_n(\mathbf{k})$ is the power spectral density for the refractive index fluctuations and $\chi(\mathbf{k})$ is a three-dimensional, normally distributed, random complex function. Since $\tilde{n}$ is a real-valued function, it implies that $\chi^*(\mathbf{k})=\chi(-\mathbf{k})$. Moreover, $\chi$ is assumed to be delta-correlated,
\begin{equation}
\langle \chi(\mathbf{k}_1) \chi^*(\mathbf{k}_2) \rangle = \Delta^3 \delta(\mathbf{k}_1-\mathbf{k}_2) .
\label{verwchi3d}
\end{equation}
There are various models for $\Phi_n(\mathbf{k})$, such as the Kolmogorov, von Karman, or Tartarskii power spectral densities \cite{scintbook}. All these power spectral densities contain a factor of the refractive index structure constant $C_n^2$. Combined with other parameters, it gives a small dimensionless quantity suitable for perturbative expansions.

Here, we do not use any specific model for the power spectral density. The expressions are left in terms of $\Phi_n(\mathbf{k})$. However, the smallness of the fluctuations is used to discard terms with more factors of $\Phi_n(\mathbf{k})$.

It follows that
\begin{align}
\langle N(\mathbf{K},z_1) N(\mathbf{K}',z_2) \rangle
= & \delta(\mathbf{K}+\mathbf{K}') \int \exp\left[-i k_z(z_1-z_2)\right] \nonumber \\
& \times \Phi_n(\mathbf{K},k_z)\ \frac{dk_z}{2\pi} .
\end{align}
where we used the fact that $\Phi_n(\mathbf{k})$ is symmetric in all its arguments. The ensemble average becomes
\begin{align}
& \langle M_0(\mathbf{K}_1,\mathbf{K}_2,z_1) M_0(\mathbf{K}_3,\mathbf{K}_4,z_2)\rangle \nonumber \\
= & k^2 \delta(\mathbf{K}_1-\mathbf{K}_2+\mathbf{K}_3-\mathbf{K}_4) \nonumber \\
& \times \exp\left[\frac{i z_1}{2k}\left(|\mathbf{K}_2|^2-|\mathbf{K}_1|^2\right)
+\frac{i z_2}{2k}\left(|\mathbf{K}_4|^2-|\mathbf{K}_3|^2\right)\right] \nonumber \\
& \times \int \exp\left[-i k_z(z_1-z_2)\right] \Phi_n(\mathbf{K}_1-\mathbf{K}_2,k_z)\ \frac{dk_z}{2\pi} .
\label{em0m0}
\end{align}
Note that $k$ is a fixed value for the wavenumber under the monochromatic approximation, whereas $k_z$ is related to the spatial Fourier transform of the medium, which has nothing to do with the frequency of the light.

Often it is assumed that the turbulent medium is delta-correlated along the propagation direction. Under this {\em Markovian approximation} the $z$-component of the power spectral density is set to zero $\Phi_n(\mathbf{K},k_z)\rightarrow\Phi_n(\mathbf{K},0)$. As a result, the integral over $k_z$ in Eq.~(\ref{em0m0}) produces a Dirac delta function in $z$. The ensemble average then simplifies, especially when two of the legs are contracted. We do not use the Markovian assumption here. Instead, we retain the non-Markovian expression in Eq.~(\ref{em0m0}).

When the single $z$-integration in Eq.~(\ref{ipewig2}) is applied to the ensemble average in Eq.~(\ref{em0m0}), it leads to the four-point vertex kernel, which we define as
\begin{align}
& \Phi_0(\mathbf{K}_1,\mathbf{K}_2,\mathbf{K}_3,\mathbf{K}_4,z,z_0) \nonumber \\
\eqtri & \int_{z_0}^z \langle M_0(\mathbf{K}_1,\mathbf{K}_2,z) M_0(\mathbf{K}_3,\mathbf{K}_4,z_1)\rangle\ dz_1 .
\end{align}
Note that the integrated $M_0$ is defined to be the second one. As a result, there are no symmetries with respect to interchanges of wave vectors in the arguments of $\Phi_0$.

\begin{widetext}
For the first two terms, we define
\begin{align}
\begin{split}
\Phi_1(\mathbf{K}_1,\mathbf{K}_4,z,z_0) \eqtri & \int \int_{z_0}^z \langle M_0(\mathbf{K}_1,\mathbf{K}',z)
M_0(\mathbf{K}',\mathbf{K}_4,z_1)\rangle\ dz_1\ \frac{d^2k'}{(2\pi)^2} \\
= & k^2 \delta(\mathbf{K}_1-\mathbf{K}_4)
\int \int_{z_0}^z \exp\left[-\frac{i (z-z_1)}{2k}\left(|\mathbf{K}_1|^2-|\mathbf{K}'|^2\right)\right] \\
& \times \int \exp\left[-i k_z(z-z_1)\right]
\Phi_n(\mathbf{K}_1-\mathbf{K}',k_z)\ \frac{dk_z}{2\pi}\ dz_1\ \frac{d^2k'}{(2\pi)^2} , \\
\Phi_1^*(\mathbf{K}_2,\mathbf{K}_3,z,z_0) \eqtri & \int \int_{z_0}^z \langle M_0(\mathbf{K}',\mathbf{K}_2,z)
M_0(\mathbf{K}_3,\mathbf{K}',z_1)\rangle\ dz_1\ \frac{d^2k'}{(2\pi)^2} \\
= & k^2 \delta(\mathbf{K}_2-\mathbf{K}_3)
\int \int_{z_0}^z \exp\left[\frac{i (z-z_1)}{2k}\left(|\mathbf{K}_2|^2-|\mathbf{K}'|^2\right)\right] \\
& \times \int \exp\left[i k_z(z-z_1)\right]
\Phi_n(\mathbf{K}_2-\mathbf{K}',k_z)\ \frac{dk_z}{2\pi}\ dz_1\ \frac{d^2k'}{(2\pi)^2} ,
\end{split}
\end{align}
where we use the symmetry $\Phi_n(\mathbf{K},k_z)=\Phi_n(-\mathbf{K},-k_z)$. Both $\Phi_1$ and $\Phi_1^*$ are symmetric with respect to an exchange of the two wave vectors.

\subsection{\label{evolz}Non-Markovian evolution equation}

The evolution equation then becomes
\begin{align}
\partial_z W_{\hat{\rho}}(z) = & -\delta_{\alpha} W_{\hat{\rho}}(z_0)\diamond \Phi_1^*(z)\diamond\alpha
-\alpha^*\diamond \Phi_1(z)\diamond\delta_{\alpha}^* W_{\hat{\rho}}(z_0) \nonumber \\
& - \int \Phi_0(\mathbf{K}_1,\mathbf{K}_2,\mathbf{K}_3,\mathbf{K}_4,z,z_0) \left[\alpha^*(\mathbf{K}_1)\alpha^*(\mathbf{K}_3)
\frac{\delta^2 W_{\hat{\rho}}(z_0)}{\delta\alpha^*(\mathbf{K}_2)\delta\alpha^*(\mathbf{K}_4)}
 +\frac{\delta^2 W_{\hat{\rho}}(z_0)}{\delta\alpha(\mathbf{K}_1)\delta\alpha(\mathbf{K}_3)}
 \alpha(\mathbf{K}_2)\alpha(\mathbf{K}_4) \right. \nonumber \\
& \left. -\alpha^*(\mathbf{K}_1)\frac{\delta^2 W_{\hat{\rho}}(z_0)}{\delta\alpha^*(\mathbf{K}_2)\delta\alpha(\mathbf{K}_3)} \alpha(\mathbf{K}_4)
-\alpha^*(\mathbf{K}_3)\frac{\delta^2 W_{\hat{\rho}}(z_0)}{\delta\alpha^*(\mathbf{K}_4)\delta\alpha(\mathbf{K}_1)}
\alpha(\mathbf{K}_2) \right]\ \frac{d^2 k_1}{(2\pi)^2}\ \frac{d^2 k_2}{(2\pi)^2}\
\frac{d^2 k_3}{(2\pi)^2}\ \frac{d^2 k_4}{(2\pi)^2} ,
\label{ipewig3}
\end{align}
\end{widetext}
where we retain the $\diamond$-contractions in the two drift terms on the bilinear vertex kernels $\Phi_1$ and $\Phi_1^*$, but due to the lack of symmetry in $\Phi_0$, we express the four dissipative terms as integrals over the wave vectors, representing contractions on the four-point vertex kernel $\Phi_0$.

The non-Markovian evolution equation for photonic states in turbulence in Eq.~(\ref{ipewig3}) is the main result. It is trace preserving: integrated over the field variable, the left-hand side gives the $z$-derivative of the trace of the state, while all the terms on the right-hand side cancel, indicating that the trace remains constant.

\section{\label{oplos}Solutions}

Due to that second-order functional derivatives on the right-hand side of Eq.~(\ref{ipewig3}), a Gaussian Wigner functional would produce polynomial factors with up to four field variables. On the left-hand side, the single $z$-derivative can only produce polynomial factors with up to two field variables from a Gaussian Wigner functional. Unless the fourth-order terms miraculously cancel among themselves on the right-hand side, this imbalance in the order of the terms on either side of the resulting equation indicates that solutions of the non-Markovian evolution equation in Eq.~(\ref{ipewig3}) cannot be in the form of Gaussian Wigner functionals. The same argument can be applied for certain non-Gaussian Wigner functionals, such as {\em polynomial Gaussian states} where the Gaussian Wigner functional is multiplied by a finite order field-variable-dependent polynomial prefactor, and {\em super-Gaussian states} where the exponent can include additional terms of arbitrary high order.

The same conclusion follows by expanding a Gaussian Wigner functional as a Taylor series (or Maclaurin series) in terms of the field variables. On the right-hand side, the four-point vertex kernel $\Phi_0$ connects factors of the exponent of the initial Gaussian state. These connections destroy the summability, so that the resulting Wigner functional loses its Gaussian state property.

Another argument for the loss of the Gaussian nature of states in turbulence follows from observing that there are no terms on the right-hand side without field variables. Assuming that the evolving state has a $z$-dependent normalization factor, as expected for a Gaussian state that becomes progressively more mixed, and there is no field-variable-independent term in its exponent, we find that the $z$-derivative of the normalization factor produces a field-variable-independent term on the left-hand side (after factoring out the state's Wigner functional) without any corresponding field-variable-independent terms on the right-hand side. Hence, the normalization factor does not evolve. It remains constant, which implies that a pure input state would remain pure. This result is unexpected for a process that involves ensemble averaging. To resolve this conundrum, we again conclude that solutions of the non-Markovian evolution equation cannot be Gaussian states.

We do not provide any definite solutions of Eq.~(\ref{ipewig3}). However, a discussion of possible solutions is in order.

\subsection{\label{oplosftra}Field transformations}

There is no physical justification to assume that the dissipative terms in the evolution equation would be suppressed relative to the drift terms, because all these terms contain the same number of $C_n^2$'s. However, it is instructive to consider the crude approximation where we discard all the dissipative terms. The resulting equation is not trace-perserving. We can make it trace preserving by adding an additional term {\em by hand}. Then it becomes
\begin{align}
\partial_z W_{\hat{\rho}}(z)
= & \kappa W_{\hat{\rho}}(z_0)-\delta_{\alpha} W_{\hat{\rho}}(z_0)\diamond \Phi_1^*(z)\diamond\alpha \nonumber \\
& -\alpha^*\diamond \Phi_1(z)\diamond\delta_{\alpha}^* W_{\hat{\rho}}(z_0) .
\label{ipetra}
\end{align}
The value of $\kappa$ can be obtained by computing the trace of this equation, which gives $\kappa=-\tr\{\Phi_1(z)+\Phi_1^*(z)\}$.

The solutions of the evolution equation in Eq.~(\ref{ipetra}) is given in terms of the initial state by a transformation of the field variables, represented as $\alpha \rightarrow Y^{\dag}(z)\diamond\alpha$ and $\alpha^* \rightarrow \alpha^*\diamond Y(z)$, where $Y(z)$ is an unknown kernel. These transformations can effect the normalization of the state. Therefore, the initial state is thus transformed to produce
\begin{align}
W_{\hat{\rho}}[\alpha^*,\alpha](z_0)
\rightarrow & \mathcal{N}(z) W_{\hat{\rho}}[\alpha^*\diamond Y^{\dag}(z),Y(z)\diamond\alpha](z_0) \nonumber \\
= & W_{\hat{\rho}}[\alpha^*,\alpha](z) ,
\end{align}
where $\mathcal{N}(z)$ provides a modification of the normalization of the state. The $z$-derivative of the $W_{\hat{\rho}}(z)$ produces
\begin{align}
\partial_z W_{\hat{\rho}}(z)
= & \left[\partial_z\mathcal{N}(z)\right] W_{\hat{\rho}}
+ \mathcal{N}(z) \alpha^*\diamond \partial_z Y(z)\diamond\frac{\delta W_{\hat{\rho}}}{\delta\alpha^*} \nonumber \\
& + \mathcal{N}(z) \frac{\delta W_{\hat{\rho}}}{\delta\alpha}\diamond\partial_z Y^{\dag}(z)\diamond\alpha .
\end{align}
The states on the right-hand side of Eq.~(\ref{ipetra}) is not affected by the transformation, being evaluated at $z_0$.

After replacing the left-hand side of Eq.~(\ref{ipetra}) by this $z$-derivative, we can extract the following equations
\begin{align}
\begin{split}
\partial_z \mathcal{N}(z) = & -\tr\{\Phi_1(z)+\Phi_1^*(z)\} , \\
\partial_z Y(z) = & -\Phi_1(z) .
\end{split}
\end{align}
In this way, we obtain a solution for the crude approximation. Since $\Phi_1(z)$ is only non-zero on the diagonal, its trace is divergent. This divergence removes divergences that appear in the other terms. Apart from being readily solvable, we do not expect this solution to be of physical significance in the scenario under investigation.

\subsection{\label{oplterm}Thermal states}

As already mentioned, the solutions of the evolution equation are not expected to be Gaussian states, even if the initial state is a Gaussian state. Nevertheless, there may be special cases where the solution can be approximated by a Gaussian state. A candidate for such a case is the thermal state. It is generically defined as
\begin{equation}
W_{\Theta}[\alpha] = \mathcal{N}_0 \det\{\Theta\}\exp(-2\alpha^*\diamond\Theta\diamond\alpha) .
\end{equation}
where $\Theta$ is a Hermitian invertible kernel that defines the thermal state. In the evolution process, we allow $\Theta$ to evolve as a function of $z$. When we substitute this state into Eq.~(\ref{ipewig3}), it becomes
\begin{widetext}
\begin{align}
\partial_z W_{\Theta}(z)
= & 2\left[\alpha^*\diamond\Theta(z_0)\diamond \Phi_1^*(z)\diamond\alpha
+\alpha^*\diamond \Phi_1(z)\diamond\Theta(z_0)\diamond\alpha\right] W_{\Theta}(z_0) \nonumber \\
& - 4\Phi_0(\mathbf{K}_1,\mathbf{K}_2,\mathbf{K}_3,\mathbf{K}_4,z,z_0)
 \left[\alpha^*(\mathbf{K}_1)\Theta(\mathbf{K}_2,\mathbf{K}_a,z_0)\alpha(\mathbf{K}_a)
\alpha^*(\mathbf{K}_3)\Theta(\mathbf{K}_4,\mathbf{K}_b,z_0)\alpha(\mathbf{K}_b) \right. \nonumber \\
& +\alpha^*(\mathbf{K}_a)\Theta(\mathbf{K}_a,\mathbf{K}_1,z_0)\alpha(\mathbf{K}_2)
\alpha^*(\mathbf{K}_b)\Theta(\mathbf{K}_b,\mathbf{K}_3,z_0)\alpha(\mathbf{K}_4) \nonumber \\
& -\alpha^*(\mathbf{K}_1)\Theta(\mathbf{K}_2,\mathbf{K}_a,z_0)\alpha(\mathbf{K}_a)
\alpha^*(\mathbf{K}_b)\Theta(\mathbf{K}_b,\mathbf{K}_3,z_0)\alpha(\mathbf{K}_4)
+\tfrac{1}{2}\alpha^*(\mathbf{K}_1)\Theta(\mathbf{K}_2,\mathbf{K}_3,z_0)\alpha(\mathbf{K}_4) \nonumber \\
& -\alpha^*(\mathbf{K}_3)\alpha^*(\mathbf{K}_a)\Theta(\mathbf{K}_a,\mathbf{K}_1,z_0)\alpha(\mathbf{K}_2)
\Theta(\mathbf{K}_4,\mathbf{K}_b,z_0)\alpha(\mathbf{K}_b) \nonumber \\
& \left. +\tfrac{1}{2}\alpha^*(\mathbf{K}_3)\Theta(\mathbf{K}_4,\mathbf{K}_1,z_0)\alpha(\mathbf{K}_2) \right] W_{\Theta}(z_0) ,
\label{thevol}
\end{align}
\end{widetext}
where repeated wave vectors are integrated over.

Since the thermal state is fully parameterized by a single kernel, we only need to find a solution for this kernel. It is done for the inverse kernel as the expectation value
\begin{equation}
\tfrac{1}{2} \Theta^{-1}(\mathbf{K}',\mathbf{K}) = \int \alpha^*(\mathbf{K})\alpha(\mathbf{K}') W_{\Theta}\ \Dcirc[\alpha] .
\end{equation}
Multiplying Eq.~(\ref{thevol}) by $\alpha^*(\mathbf{K}_a)\alpha(\mathbf{K}_b)$ and integrating over $\alpha$, we obtain an evolution equation for this expectation value in terms of higher order expectation values. All these expectation values are obtained with the aid of a generating functional, given by
\begin{align}
\mathcal{W}_{\Theta}[\nu,\mu^*] = & \int W_{\Theta}
\exp\left(\alpha^*\diamond\nu+\mu^*\diamond\alpha\right)\ \Dcirc[\alpha] \nonumber \\
= & \exp\left[\tfrac{1}{2}\mu^*\diamond \Theta^{-1}\diamond\nu\right] .
\end{align}
The uneven expectation values vanish. Higher order expectation values are all expressed in terms of $\Theta^{-1}$. After evaluating the even expectation values in the equation, many of the terms cancel. The resulting equation then reads
\begin{widetext}
\begin{align}
\partial_z \Theta^{-1}(\mathbf{K}_b,\mathbf{K}_a,z)
= & \Theta^{-1}(\mathbf{K}_x,\mathbf{K}_y,z_0)\Phi_0(\mathbf{K}_y,\mathbf{K}_a,\mathbf{K}_b,\mathbf{K}_x,z,z_0)
+\Theta^{-1}(\mathbf{K}_x,\mathbf{K}_y,z_0)\Phi_0(\mathbf{K}_b,\mathbf{K}_x,\mathbf{K}_y,\mathbf{K}_a,z,z_0) \nonumber \\
& -\Theta^{-1}(\mathbf{K}_b,\mathbf{K}_0,z_0)\Phi_1^*(\mathbf{K}_0,\mathbf{K}_a,z)
-\Phi_1(\mathbf{K}_b,\mathbf{K}_0,z)\Theta^{-1}(\mathbf{K}_0,\mathbf{K}_a,z_0) .
\end{align}
\end{widetext}
with repeated wave vectors being integrated over. Thus we obtained an expression for the evolving inverse kernel of the thermal state due to scintillation process that depends linearly on its initial inverse kernel.

\subsection{\label{verliesopl}Loss-based model}

A model that has been proposed for the evolution of a multiphoton state in turbulence \cite{semenov} is based on modeling the scintillation process as a loss mechanism. The state is defined in terms of a $P$-distribution. A Wigner functional can be represented in terms of a $P$-functional by the functional convolusion integral
\begin{equation}
W[\alpha] = \mathcal{N}_0 \int P[\alpha'] \exp\left(-2\|\alpha-\alpha'\|^2\right)\ \Dcirc[\alpha'] .
\end{equation}
To introduce the effect of photon loss, we apply a scaling factor in the argument of the $P$-functional, while maintaining its normalization. The result is
\begin{equation}
W[\alpha] = \frac{\mathcal{N}_0}{L^{\Omega}} \int P\left[\frac{\alpha'}{L}\right]
\exp\left(-2\|\alpha-\alpha'\|^2\right)\ \Dcirc[\alpha'] ,
\end{equation}
where $0<L<1$ is the loss, and $\Omega$ is the cardinality of the functional phase space.

As an example, we consider a coherent state, given by
\begin{equation}
P[\alpha'] = (2\pi)^{\Omega} \delta[\alpha'-\zeta] ,
\end{equation}
where $\zeta$ is the parameter function of the coherent state. It leads to
\begin{align}
W[\alpha] = & \frac{\mathcal{N}_0}{L^{\Omega}} \int \delta\left[\frac{\alpha'}{L}-\zeta\right]
\exp\left(-2\|\alpha-\alpha'\|^2\right)\ \mathcal{D}[\alpha'] \nonumber \\
= & \mathcal{N}_0 \exp\left(-2\|\alpha-L\zeta\|^2\right) ,
\end{align}
where $\mathcal{D}[\alpha']=(2\pi)^{\Omega}\ \Dcirc[\alpha']$. We see that the loss reduces the amplitude of the parameter function, as expected. However, we would also expect the scintillation to cause a distortion of the spatial mode, represented by $\zeta$. Statistically, distortions produce broader modes.

As a generalization, we introduce a probability distribution $f_L(L,z)$ that depends on $z$ for the loss, so that
\begin{align}
W[\alpha](z) = & \int_0^1 f_L(L,z) \frac{\mathcal{N}_0}{L^{\Omega}} \int P\left[\frac{\alpha'}{L}\right] \nonumber \\
& \times \exp\left(-2\|\alpha-\alpha'\|^2\right)\ \Dcirc[\alpha']\ dL \nonumber \\
= & \mathcal{N}_0 \int P[\alpha'] \int_0^1 f_L(L,z) \nonumber \\
& \times \exp\left(-2\|\alpha-\alpha'L\|^2\right)\ dL\ \Dcirc[\alpha'] .
\label{semenov}
\end{align}
In the last expression, we redefined the integration field variable $\alpha'\rightarrow\alpha'L$ to remove $L$ from $P[\alpha']$. The result is a mixed state that lost its Gaussian nature. The initial probability distribution $f_L(L,z_0)$ imposes $L=1$.

\begin{widetext}
When we substitute Eq.~(\ref{semenov}) into the evolution equation, it becomes
\begin{align}
\partial_z W_{\hat{\rho}}(z) = & \mathcal{N}_0 \int P[\alpha'] \int_0^1 \partial_z f_L(L,z)
\exp\left(-2\|\alpha-\alpha'L\|^2\right)\ dL\ \Dcirc[\alpha'] \nonumber \\
= & -2\mathcal{N}_0 \int P[\alpha']
\left({\alpha'}^*\diamond\Phi_1^*(z)\diamond\alpha+\alpha^*\diamond\Phi_1(z)\diamond\alpha'\right)
\exp\left(-2\|\alpha-\alpha'\|^2\right)\ \Dcirc[\alpha'] \nonumber \\
& - 4 \mathcal{N}_0 \int P[\alpha']\exp\left(-2\|\alpha-\alpha'\|^2\right)
\Phi_0(\mathbf{K}_1,\mathbf{K}_2,\mathbf{K}_3,\mathbf{K}_4,z,z_0) \nonumber \\
& \times \left[\alpha^*(\mathbf{K}_1)\alpha'(\mathbf{K}_2)-{\alpha'}^*(\mathbf{K}_1)\alpha(\mathbf{K}_2)\right]
\left[\alpha^*(\mathbf{K}_3)\alpha'(\mathbf{K}_4)-{\alpha'}^*(\mathbf{K}_3)\alpha(\mathbf{K}_4)\right]\ \Dcirc[\alpha'] .
\end{align}
where repeated wave vectors are integrated over. The second order functional derivatives produce identities contracted on $\Phi_0$ leading to $\Phi_1$'s, which cancel the $\Phi_1$-terms without $\alpha'$. The remaining fourth-order terms factorize. Since both sides of the equation contain the same arbitrary initial state $P[\alpha']$, the parts inside the functional integrals without these $P$-functionals must be equal:
\begin{align}
\int_0^1 \partial_z f_L(L,z) \exp\left(-2\|\alpha-\alpha'L\|^2\right)\ dL
= & -2 \exp\left(-2\|\alpha-\alpha'\|^2\right) \left({\alpha'}^*\diamond\Phi_1^*(z)\diamond\alpha+\alpha^*\diamond\Phi_1(z)\diamond\alpha'\right) \nonumber \\
& - 4 \exp\left(-2\|\alpha-\alpha'\|^2\right)
\left[\alpha^*(\mathbf{K}_1)\alpha'(\mathbf{K}_2)-{\alpha'}^*(\mathbf{K}_1)\alpha(\mathbf{K}_2)\right] \nonumber \\
& \times \left[\alpha^*(\mathbf{K}_3)\alpha'(\mathbf{K}_4)-{\alpha'}^*(\mathbf{K}_3)\alpha(\mathbf{K}_4)\right] \Phi_0(\mathbf{K}_1,\mathbf{K}_2,\mathbf{K}_3,\mathbf{K}_4,z,z_0) .
\label{ipeloss}
\end{align}
\end{widetext}
The only unknown in the equation is $f_L(L,z)$. It thus seems to represent an equation with which $f_L(L,z)$ can be solved. The resulting equation is trace preserving, which can be shown by evaluating the trace over $\alpha$.

Unfortunately, closer inspection of this equation shows that it is not a valid equation. Expansions on either side in terms of the field variables produce terms on the left-hand side that cannot be matched by equivalent terms on the right-hand side.

Moreover, the left-hand side does not contain any spatial information, while the spatial information exists in terms of the scintillation kernels on the right-hand side. To make this observation more explicit, we use the left-hand side of the expression as a generating functional for the moments of $L$. As a result, one can obtain equations for all the moments of $f_L(L,z)$. However, the resulting equations for these moments cannot be solved, because of an imbalance of wave vector dependences on either side of such equations.

To demonstrate this imbalance, we consider the equation for the first moment, given by
\begin{equation}
M_1(z) = \int_0^1 f_L(L,z) L\ dL .
\end{equation}
It is obtained by computing the functional derivatives with respect to $\alpha'$ and $\alpha^*$, and then setting all field variables to zero. The result reads
\begin{equation}
2 \delta(\mathbf{K}_a-\mathbf{K}_b) \partial_z M_1(z) = -2 \Phi_1(\mathbf{K}_a,\mathbf{K}_b,z) .
\end{equation}
To remove the Dirac delta function, we integrate over one of the transverse wave vectors. The left-hand side becomes independent of the wave vectors, but the right-hand side retains a wave vector dependence due to the diagonal dependence of $\Phi_1$. (This wave-vector dependence would disappear in the Markovian limit.) Higher order moments produce ambiguous equations, because there are different sets of functional derivatives that produce the same higher order moment, but with different terms on the right-hand side.

In the end, we conclude that the loss model does not provide a valid solution for the evolution equation, especially not in the non-Markovian case. The reason is the lack of spatial information in the model.

\section{\label{conclu}Conclusions}

A derivation is provided of the non-Markovian evolution equation for multiphoton states propagating through turbulence. The Wigner functional approach used, leads to a Fokker-Planck equation for the Wigner functional of the state. It contains drift terms with two-point kernels denoted by $\Phi_1$'s and dissipative (or diffusion) terms with four-point vertex kernels denoted by $\Phi_0$.

The form of the evolution equation leads us to conclude that its solutions do not include Gaussian Wigner functionals. Neither do they include polynomial Gaussians or super-Gaussians. As a result, exact solutions are difficult to find.

It is instructive to see that, without the dissipative terms, which contain the four-point vertex kernel, the equation can be solved in terms of transformations of the arguments of the initial Wigner functional. It would allow such an initial Wigner functional to retain its Gaussian nature. However, it is not reasonable that the disspative terms can be discarded relative to the drift terms.

We also consider the possibility that some states can be approximated by a Gaussian state. For this purpose, we consider a thermal state parametrized by a single bilinear kernel. Computing the second moment of the evolution equation, we obtain an evolution equation for this kernel.

Finally, we demonstrate that the state based on modeling the scintillation process as a simple loss process does not provide a solution for the non-Markovian evolution equation. The reason is that this model does not treat the spatiotemporal degrees of freedom in a way that is suitable for a solution of this equation.

\section*{Acknowledgements}

This work was funded by the South African Quantum Technology Initiative (SA QuTI) through the Department of Science and Innovation of South Africa.


\end{document}